\documentclass[12pt]{article}
\usepackage{hyperref}
\usepackage{lscape}
\usepackage{amsmath}
\usepackage{amssymb}

\renewcommand{\title}[1]{%
    \bigskip%
    \begin{center}%
    \Large\bf #1%
    \end{center}%
    \vskip .2in}

\renewcommand{\author}[1]{%
    {\begin{center}
    #1
    \end{center}}}
\newcommand{\address}[1]{\vspace{-1.7em}\vspace{0pt}
    {\begin{center}
    \it #1
    \end{center}}}

\begin{document}


\title{Milne boost from galilean gauge theory}

\author
{
Rabin Banerjee  $\,^{\rm a,b}$,
Pradip Mukherjee $\,^{\rm c,d}$}
\address{$^{\rm a}$S. N. Bose National Centre 
for Basic Sciences, JD Block, Sector III, Salt Lake City, Kolkata -700 098, India }

\address{$^{\rm c}$Department of Physics, Barasat Government College,\\10, KNC Road, Barasat, Kolkata 700124, India.

 }

\address{$^{\rm b}$\tt rabin@bose.res.in}
\address{$^{\rm d}$\tt mukhpradip@gmail.com}

\begin{abstract}Physical origin of Milne boost
 invariance of the Newton Cartan spacetime is traced to the effect of local Galilean boosts in its metric structure, using galilean gauge theory. Specifically, we do not require any gauge field to understand Milne boost
 invariance.
\end{abstract}

{\section{Introduction}}
 Newton Cartan (NC) geometry \cite{Cartan-1923,Cartan-1924,Havas, ANDE,TrautA,Kuch,Daut,EHL, MALA} was introduced by Cartan \cite{Cartan-1923,Cartan-1924} to provide a geometric formulation of Newtonian gravity. Interest in this field was rekindled when non relativistic spatial diffeomorphism was shown to be an important tool in the study of unitary Fermi gas \cite{SW}, which has
applications in the study of 
  fractional quantum Hall effect (FQHE) \cite{SW, HS,M1,M2,GS,Wu}. The NC geometry possesses a characteristic invariance 
  under the Milne boosts \cite{K}, in addition to the diffeomorphism invariances \cite{DH}. The symmetry group is abelian \cite{DH} which has inspired many researchers to introduce a gauge field in the NC structure \cite{JEN, JK, OBN} to represent this symmetry. However, why such a gauge field will be mixed up with the geometric elements of NC manifold
  is a pertinent question. To answer such a question the redundancy of the metric NC spacetime, due to the Milne boosts, has been utilised,. The idea of a gauge field as part of a geometry may be  novel but to some extent uneasy. Different approaches to the Milne boost symmetry are, thus, essential.  

  Infact, the concept of  nonrelativistic diffeomorphism invariance (NRDI), which led to the recent understanding of the NC spacetime, was marred with several problems, of which passage to Galilean symmetry in the flat limit is a representative one. One finds \cite{SW} that a gauge field must always be invoked and in the flat limit the corresponding gauge parameter must be equated with the Galilean boost parameter. This unexpected connection, though compatible with a gauge field in the NC geometry, leads to  new questions, concerning the   coupling of a free Schrodinger theory in the absence of any gauge field \cite{BGM}.  In this context it may be remembered that the coupling of the non relativistic theories with curved background can be consistently done 
\cite{BGM, BMM1, BM4} where no preexisting gauge field was needed.  
 Also, the NC covariant form of the Schrodinger field (with or without gauge interaction) was obtained by taking $c\to \infty$ limit of an appropriate relativistic theory \cite{BGM}. 
 
 Presumably, the approach of introducing a  gauge field in the NC structure was  triggered by \cite{SW}. The rational behind this was the presence of Milne boosts which demonstrate certain redundancy in the metric structure. However,  this remedy has several side effects. The most important one is that the symmetric connection ceases to be Milne invariant if it is to be gauge invariant and conversely \cite{JK}. Also remember that such a gauge field was never required in the earlier works, many of them by past masters \cite{Cartan-1923,Cartan-1924,Havas, ANDE,TrautA,Kuch,Daut,EHL, MALA, DH}.
 
 We show in the present paper that the invariance under Milne boosts can be completely explained within the confines of the usual metric structure. 
 In what follows we will apply
a systematic algorithm for obtaining non relativistic diffeomorphism invariance by localising (gauging) a Galileo invariant theory, 
called galilean gauge theory (GGT)
\cite{BMM2, BM4, BMM1, BMM3, AM}.
 The outcome of GGT is the coupled form of a nonrelativistic  theory  with curved background.
 The identity of the curved background has been  ascertained by deriving the metric formulation of the NC geometry from the first order GGT \cite{BMM2, BM4,BMM3}.Naturally one expects that new vistas in the understanding of  Milne boosts may be opened up from GGT. In this paper we will show that  this hope is not belied. Specifically, the Milne boosts will be shown to be the remnants of the local galilean boost symmetry of the vierbein formulation of the NC space time.
 
The study of Milne boost invariance has its intrinsic interest also.
 The NC geometry is the non-relativistic limit of Riemann-Cartan geometry. But the Milne boost symmtry has no relativistic analog.  The authors of \cite{DH} relate the arbitrariness of the symmetric connection which is manifestated by an arbitrary two form, with the Milne boost invariance.  In our approach the symmetric connection is obtained naturally and its invariance follows as a consequence of the transformations of the basic fields. Also, we show that the complete Milne transformations are  a consequence of local galilean boosts reflected in the metric structure. This happens because the galilean boosts affect space and time asymmetrically, unlike the Lorentz transformations. 
 


The organisation of the paper is as follows, In section 2 we give a brief review of Milne boost trasformations in metric NC theory. In the next section the connection between the vielbein formulation of GGT and the metric structure of the NC spacetime is recapitulated, This is necessary to set the stage for further discussion and to introduce the symbols. Sections 4 and 5 contain the new results of the present work.
The paper ends with  concluding remarks in section 6.
{\section{The metric Newton-Cartan theory and Milne boosts}} The Newton-Cartan (NC) theory, in the standard form, is formulated as a metric theory. The special role of time in the galilean models is manifested in the lack of a single nondegenerate metric in space time. Rather there are two degenerate metrics which are given by a temporal one form $\tau_\mu$ and a degenerate second rank metric $h^{\mu\nu}$ so that
\begin{eqnarray}
h^{\mu\nu}\tau_\nu = 0\label{NCnew}
\end{eqnarray} 
The quotient space $M/ker(\tau_\mu)$ is one dimensional and points the direction in which time flows. A vector $v^\mu$ corresponds to the metric one form $\tau_\mu$ such that,
\begin{eqnarray}
v^{\mu}\tau_{\mu} = 1\label{NCnew1}
\end{eqnarray}
and a second rank covariant tensor $h_{\mu\nu}$ is introduced to project any vector to a space like one form. The appropriate projection operator is,
\begin{eqnarray}
h_{\mu\nu} 
h^{\nu\rho} = \delta_\mu^\rho - \tau_\mu v^\rho
 = P^{\mu}_\rho\label{NCnew3}
\end{eqnarray}

Another piece of the scenario is the connection. The expression of the symmetric and metric compatible connection goes as
\begin{eqnarray}
\Gamma^{\rho}_{\mu\nu} = v^\rho\partial_{\left(\mu\right.}\tau_{\left.\nu\right)} +\frac{1}{2}h^{\rho\lambda}
                          \left(\partial_{\left(\mu\right.}h_{\left.\nu\right)\lambda} - \partial_\lambda h_{\mu\nu}\right)
                          + h^{\rho\lambda}K_{\lambda\left(\mu\right.}\tau_{\left.\nu\right)}\label{D}
\end{eqnarray}
where, $K_{\mu\nu}$ is an arbitrary two form. Note that the connection is not uniquely determined by the metrics. 
This is unlike the pseudo Riemannian case where the symmetric, metric compatible connection is unique.

The Milne boosts are defined by \cite{DH1},
\begin{eqnarray}
\tau_\mu\to\tau^\prime_\mu & = &\tau_\mu; \hskip.3cm h^{\mu\nu} \to h^{\prime{\mu\nu}}=h^{\mu\nu}\label{n11}\\
v^\mu  \to v^{\prime\mu} & = & v^\mu + h^{\mu\nu}\psi_\nu\label{n1}\\
h_{\mu\nu} \to h^\prime_{\mu\nu} & = & h_{\mu\nu} - (\tau_\mu P_\nu^\rho + \tau_\nu P_\mu^\rho)\psi_{\rho} + \tau_\mu\tau_\nu h^{\rho\sigma}\psi_\rho\psi_\sigma\label{n2}
\end
{eqnarray}
where $\psi_\nu$ is any arbitrary one form. Here $\tau_{\mu}$ and $h^{\mu\nu}$ are kept unchanged. It is simple to check that the NC relations (\ref{NCnew}, \ref{NCnew1}, \ref{NCnew3}) are invariant under the Milne boosts (\ref{n11},\ref{n1}, \ref{n2}).

We then come to the transformation of the connection (\ref{D}).
 The last term of the expression for the connection contains an arbitrary two form $K_{\mu\nu}$.
 At most we can write
\begin{equation}
K^\prime{}_{\mu\nu} =K_{\mu\nu} - \partial_{\left[\mu\right.} \phi_{\left.\mu\right]}\label{KM}
\end{equation}
with the one form $\phi_\mu$ arbitrary. Its transformation can be fixed by demanding 
\begin{equation}
\Gamma^{\prime\rho}_{\mu\nu} = \Gamma^{\rho}_{\mu\nu}
\end{equation}
under Milne transformations. Then an explicit ansatz for $\phi_\mu$ is obtained   \cite{DH},
\begin{equation}
\phi_\mu = \psi_\mu -\left(\psi_\lambda v^\lambda + \frac{1}{2}h^{\lambda\nu}\psi_\lambda\psi_\nu\right)\tau_\mu\label{phi}
\end{equation}
 We, on the other hand, are able to provide a deduction
of the full connection in terms of the first order variables, namely, the vierbeins and the spin connections. This will enable us to calculate the variation of the connection from first principles. We find that the connection is invariant.

At this point it can be easily seen why the invariance of the connection
 is lost when we exploit the arbitrariness of $K_{\mu\nu}$ to introduce a gauge field $A_\mu$. The arbitrary  
 two form is then expressed as
$K_{\mu\nu}= \partial_\mu A_\nu - \partial_\nu A_\mu $. The transformation of  $A_\mu $ is fixed beforehand and the corresponding transformation of $K_{\mu\nu}$ is such that the connection is no longer invariant \cite{JEN, JK}.

The Milne boost transformations given by (\ref{n11},\ref{n1}, \ref{n2}) preserve the NC algebra. We find that the set $({\tau_\mu} ,h^{\mu\nu})$ remains invariant while $(v^\mu, h_{\mu\nu})$ changes. Alternatively, it has been formulated in a complimentary manner \cite{JEN} where $v^\mu$ and $h_{\mu\nu}$ are invariant whereas $\tau_\mu$ and $h^{\mu\nu}$ transform  appropriately , so as to keep the 
NC algebra intact. Using the reduction of relativistic to nonrelativistic field models \cite {JEN, JK}, however,  the first set where $({\tau_\mu} ,h^{\mu\nu})$ does not change was found. Our method leads to a unique form that matches with the one where $({\tau_\mu} ,h^{\mu\nu})$ are invariant, as will be seen in the following.
\bigskip
{\section{Galilean gauge theory and Newton Cartan spacetime} Galilean gauge theory as well as its connection with NC geometric elements, with \cite{RBPMt} or without torsion, have been developed  \cite{BMM2, BMM3, BM4}
 over the last few years. We provide
an illustration of the algorithm 
using the free Schrodinger field action
\begin{equation}
S = \int dt  \int d^3x  \left[ \frac{i}{2}\left( \psi^{*}\partial_{0}\psi-
\psi\partial_0\psi^{*}\right) -\frac{1}{2m}\partial_k
\psi^{*}\partial_k
\psi\right].
\label{globalaction} 
\end{equation}
which is invariant under the global Galilean transformations,
 \begin{equation}
  x^\mu \to x^\mu + \xi^\mu\label{globalgalileans}
\end{equation}
where $\xi^\mu$ is given by $\xi^{0} =-\epsilon$ and $\xi^{i} = \eta^{i}-u^{i}t $ with
$\eta^i=\epsilon^{i}+ \lambda^{i}{}_{j}x^{j}$
 The constant parameters $\epsilon$, $\epsilon^{i}$, $\lambda^{ij}$ and $u^{i}$, respectively, represent time and space translation, spatial rotations and galilean boosts.
 $\lambda^{ij}$ are antisymmetric under interchange of the indices. 
 If these parameters are elevated to functions of space and time, the corresponding symmetry will involve local Galilean transformations. To capture the distinction of time and space, the time translation parameter is generalised to be a function of time only.
 
  The first step is to write the corresponding locally Galilean symmetric model in terms of local coordinates. One can readily write this, following the algorithm given in \cite{BMM1, BMM3, BM4}, as
\begin{equation}
S= \int dx^0d^3x\det{{\Sigma_\alpha}^{\mu}}\left[\frac{i}{2}(\psi^{*}\nabla_{0}\psi
-\psi\nabla_{{0}}\psi^{*}) -\frac{1}{2m}\nabla_a\psi^{*}
\nabla_a\psi
\right]
\label{localschaction} 
\end{equation} 
 The time coordinate in the local system will be denoted by ${0}$ and the space coordinates by $a$. Collectively, the local coordinates will be denoted by the initial letters of the Greek alphabet (i.e. $\alpha,\beta$ etc.). The action (\ref{localschaction}) has been geometrically interpreted as a diffeomorphism invariant theory in the NC space time. ${\Sigma_\alpha}^{\mu}$ are the vierbeins with ${\Lambda_\mu}^{\alpha}$ as its inverse.
 The local covariant derivatives are
 \begin{eqnarray}
\nabla_{{0}}\psi &=&{\Sigma_{{0}}}^0D_0 \psi+{\Sigma_{{0}}}^k D_k\psi
\nonumber\\
\nabla_a\psi &=&{\Sigma_a}^{k}D_k\psi.
\label{nab1}
\end{eqnarray}
where
\begin{eqnarray}
D_\mu\psi=\partial_\mu\psi+iB_\mu\psi \label{firstcov1}
\end{eqnarray}
are the covariant drivatives in the global coordinates.
 Also, in the adapted coordinates, $\Sigma_a{}^0 = 0$. Hence
\begin{eqnarray}
 \det{{\Sigma_\alpha}^{\mu}} = \frac{\det(\Sigma_a{}^k)^{-1}}{\Sigma_0{}^0}
\end{eqnarray}

The new fields $B_\mu$
have the structures,
\begin{eqnarray}
B_\mu = \frac{1}{2}{B_\mu}^{ab}\omega_{ab} + {B_\mu}^{a{0}}\omega_{a}
\label{gaugefields}
\end{eqnarray}
where $\omega_{ab}$ and $\omega_{a}$ are respectively the generators of rotations and Galileo boosts. 
The  transformations of  $\Sigma_\alpha{}^\mu$ and $B_\mu$ that preserve the invariance of (\ref{localschaction}) are given by \cite{BM4},
\begin{align}
\delta_0 {\Sigma_0}^{0} &= -\xi^\nu {\partial_\nu{\Sigma}_0}^{0}+ {\Sigma_0}^{\nu}\partial_{\nu}\xi^{0} \nonumber\\
\delta_0 {\Sigma_0}^{k} &= -\xi^\nu {\partial_\nu{\Sigma}_0}^{k}+ {\Sigma_0}^{\nu}\partial_{\nu}\xi^{k} + u^b{\Sigma_b}^{k}\nonumber\\
\delta_0 {\Sigma_a}^{k} &= -\xi^\nu {\partial_\nu{\Sigma}_a}^{k}+ {\Sigma_a}^{\nu}\partial_{\nu}\xi^{k} -\lambda_a{}^b{\Sigma_b}^{k}\nonumber\\
\delta_0 {\Lambda_0}^{0} &= -\xi^\nu {\partial_\nu{\Lambda}_0}^{0}+ {\Lambda_\nu}^{0}\partial_{0}\xi^{\nu}\nonumber\\
\delta_0 {\Lambda_0}^{a} &= -\xi^\nu {\partial_\nu{\Lambda}_0}^{a}+ {\Lambda_\nu}^{a}\partial_{0}\xi^{\nu} - u^a{\Lambda_0}^{0}\nonumber\\
\delta_0 {\Lambda
_k}^{a} &= -\xi^\nu {\partial_\nu{\Lambda}_k}^{a}+ {\Lambda_{\nu}}^a\partial_{k}\xi^{\nu} -\omega_c{}^a{
\Lambda
_k}^{c}\nonumber\\
\delta_0 {B_k} &= -\xi^\nu \partial_\nu{B_k}- {B_i}\partial_{k}\xi^{i} + m\partial_k(u^ix_i)-mu^b{\lambda_k}^{b}\nonumber\\
\delta_0 {B}_{0} &= -\xi^\nu \partial_\nu B_0 + \partial_{0}\xi^{\mu}B_\mu + m \dot{u}_ix^i+ mu^b{\Lambda_k}^{b}{\Lambda_0}^{0}{\Sigma}_{0}^{k} 
\label{delth3}
\end{align}

The spatial part $\Lambda_k{}^a$ is the inverse of $\Sigma_a{}^k$. 
The transformations of $\Lambda_{\nu}{}^{\alpha}$ of course follows from above as $\Lambda$ is inverse of $\Sigma$
The transformations (\ref{delth3}) show that that the lower index (of $\Sigma_\alpha{}^\mu$ ,for instance)
transform under the local transformations, whereas the upper index transform under diffeomorphism,
$x^\mu \to x^\mu + \xi^\mu$. Thus the theory (\ref{localschaction}) has two types of symmetry, diffeomorphisms and local galilean transformations.

 It has been proved that the 4-dim spacetime obtained above is the NC
manifold. This is done by showing that the metric formulation of our theory contains the same structures and satisfy the same structural relations as in NC space - time \cite{BMM2}.
 We begin with the definitions of the  metric, 
 \begin{equation}
h^{\mu\nu}={\Sigma_a}^{\mu}{\Sigma_a}^{\nu}
\label{spm}
\end{equation}
and the one form
\begin{equation}
\tau_{\mu}={\Lambda_\mu}^{0}
\label{tem}
\end{equation}
Since $\Sigma_a{}^0 = 0$,
\begin{eqnarray}
h^{\mu 0} = \Sigma_a{}^\mu\Sigma_a{}^0 = h^{0\mu} =0\label{me}
\end{eqnarray}
 Using the transformations (\ref{delth3}), we can show that both $h^{\mu\nu}$ and ${\tau}_\mu$ satisfy appropriate trnsformations under diffeomorhism of the manifold. Note that as a consequence of $\Sigma_a{}^0 =0$ in our coordinates, $\Lambda_k{}^0 = 0$. So $\tau_\mu = (1,0,0,0) $.

>From the above definitions we get
\begin{equation}
h^{\mu\nu}\tau_{\nu} = 0
\label{deg}
\end{equation}
 We can define a timelike vector $v^\mu$
as 
\begin{equation}
v^{\mu}=
{\Sigma_0}^{\mu}\hspace{.3cm};\hspace{.3cm}\tau_{\mu}v^{\mu} = 1
\label{tem1}
\end{equation}
and the covariant tensor
\begin{equation}
h_{\nu\rho}=\Lambda_{\nu}{}^{a} \Lambda_{\rho}{}^{a}
\label{spm2}
\end{equation}
Clearly,
\begin{align}
h_{\mu\nu}v^\nu &=  {\Lambda_{\mu}}^a {\Lambda_\nu}^a {\Sigma_0}^\nu\notag\\ &=  {\Lambda_{\mu}}^a\delta_0^a\notag\\ &=0\label{n}
\end{align}
Finally 
\begin{equation}
h^{\mu\lambda}h_{\lambda\nu} = P^\mu_\nu = \delta^\mu_\nu - v^\mu {\tau}_{\nu}\label{spm3}
\end{equation}
Thus
$h^{\mu\nu}$,
 $\tau_\nu$ define the metrics
 of the NC geometry.
 
 If we use the vierbein postulate, the affine connection is obtained in the first order variables. Symmetrizing this connection
and using our identifications we have shown earlier that the standard Dautcourt form (\ref{D})is obtained where the nonunique last term is shown to be
\begin{eqnarray}
 h^{\rho\lambda}K_{\lambda\left(\mu\right.}\tau_{\left.\nu\right)} = {\Sigma_a}^\rho\left(B_{\left(\mu\right.}^{a0}\tau_{\left.\nu\right)}   \right)\label{extra}
\end{eqnarray}
So we have conclusively proved that the curved spacetime indeed has the NC geometry. Thus using 
 GGT one can couple a non relativistic theory with background NC spacetime. The particular benefit is to directly arrive at the convenient galilean coordinates. A great merit of the GGT is apparent now. The coupling of a field theory with the NC background is known to be very difficult because it has to be such that the fields properly transform in the adapted coordinates. GGT chooses this frame inherently. 
\bigskip
{\section{Local boost symmetry}} In this section we will discuss about the effect of local galilean boosts in the metric formulation of NC gravity. In the relativistic case the local Lorentz transformations do not affect the metric or the symmetric metric compatible connection. The situation is otherwise in the non relativistic analog. The exact way in which this phenonenon is connected with the degeneracy of the metric structure will be clear from the following analysis. Also we will show in this and the following section that the effect of galilean boost percolates through the torsionless metric theory in the form of Milne boost. The spin connection associated with the boost sector explains the arbitrariness of the metric compatible connection. A definite expression for the arbitrary part of the NC connection, which was derived earlier \cite{BMM2}, sheds new light on the relation between the local galilean boosts and the two form $K_{\mu\nu}$. 
Finally, the result that Milne boosts are nothing but space time transformations rules out the necessity of associating a gauge field to  it, as has been done in the literature \cite{JEN, JK, OBN}.
{\subsection {Residual effect of the local galilean transformation in the metric formulation of NC sacetime}} Let us consider the NC geometry in the first order formulation of GGT. This is defined by the tetrad $\Sigma_\alpha{}^\mu$ and the spin connection $B_\mu$, the most general transformations of which under diffeomorhism and local Galilean transformations are given by
 (\ref{delth3}).
Let us now specialize these transformations when only local Galilean boost is given. This is done easily by putting $\xi^\mu = 0$ and $\omega^{ab} = 0$ in  (\ref{delth3}).

 Now \footnote{Note that we are considering infitesmal boost parameter}
\begin{eqnarray}
\delta_0 {\Sigma_0}^{k} &=& \Sigma_b{}^k u^b ;\hskip .5cm \delta_0 {\Sigma_a}^{k} = 0\label{n3}\\
\delta_0 {\Lambda_0}^{a} &=&  u^a{\Lambda_0}^{0};\hskip .5cm \delta_0 {\Lambda_k}^{a} = 0\label{n4}
\end{eqnarray}
These relations may be written in a covariant form. Specifically,
\begin{equation}
\delta_0 \Lambda_{\mu}{}^a=\delta_{\mu}^0u^a\Lambda_0{}^0
=\delta_{\mu}^{\nu}u^a\Lambda_{\nu}{}^0=u^a\Lambda_{\mu}{}^0\label{W}
\end{equation}
where $\Lambda_{k}{}^0=0$ has been used. Using (\ref{tem}) and (\ref{n4}), we get, 
\begin{equation}
\delta_0 \tau_\mu = \delta_0 {\Lambda_\mu}^0 = 0 \label{nt}
\end{equation}
The case of $h^{\mu\nu}$ is similar,
\begin{equation}
\delta_0 h^{\mu\nu} = \delta_0 ({\Sigma^\mu}_a{\Sigma^\nu}_a) = 0\label{nh} 
\end{equation}
where use has been made of equation (\ref{n3}). 

We observe that under the local boost transformation, the two degenerate metrics $\tau_\mu$
and $h^{\mu\nu}$ remain invariant. The changes occur to the auxiliary set $(v^\mu , h_{\mu\nu})$. The alternative way of writing the Milne transformations, where $(v^\mu, h_{\mu\nu})$ are invariant but $\tau_\mu, h^{\mu\nu}$ change, though algebraically feasible, is not dynamically suggested.
Also, note that should there have been a nondegenerate metric, as happens for Riemannian geometry, the auxiliary structures would not exist.
Then there would be no Milne type nontrivial transformation. This explains why such symmetry is not generated by three dimensional rotation in space, in the NC spacetime.
\bigskip
\subsection{The connection} So far we have not considered what happens to the connection under local galilean transformation. The affine connection $\Gamma_{\nu\mu}^{\rho}$ in the first order form is easy to abstract from the vielbein postulate,
 \begin{equation}
\nabla_\mu{\Lambda^\alpha}_{\nu} = \partial_{\mu}{\Lambda^\alpha}_{\nu} - \Gamma_{\nu\mu}^{\rho}{\Lambda^\alpha}_{\rho}
+B^{\alpha}{}_{\mu\beta}{\Lambda^\beta}_{\nu} =0 
 \label{P}
\end{equation}
and is given by
 \begin{eqnarray}
\Gamma_{\nu\mu}^{\rho} &= \partial_{\mu}{\Lambda_{\nu}}^\alpha {\Sigma_\alpha}^{\rho}
+B^{\alpha}{}_{\mu\beta}{\Lambda_{\nu}}^\beta
{\Sigma_\alpha}^{\rho}\label{con1}
\end{eqnarray}
where the spin connection $B^{\alpha}{}_{\mu\beta}$ appears in addition to the vierbeins.
 The generator of the galilean boost
is $\omega^{a} = mx^a$. From (\ref{delth3}), specialising for local galilean boost, we get,
\begin{eqnarray}
\delta_0B_k
            = m\partial_k(u^ax_a) - m  u^b\Lambda_k{}^b\nonumber\\
            = m\partial_ku^ax^a
\end{eqnarray}
Similarly, 
\begin{eqnarray}
\delta_0B_0
            = m\partial_0(u^ax_a) - m  u^b\Lambda_0{}^b\nonumber\\
            = m\partial_0u^ax^a
\end{eqnarray}
where use has been made of the fact that in the adapted coordinates $u^0$ is zero in both the basis.
Simplifying, we get, 
\begin{equation}
\delta_0B_\mu{}^{a0} 
                                       = \partial_\mu u^a
\end{equation}
where (\ref{gaugefields}) has been used.

Now we can find the cherished variation of $\Gamma_{\nu\mu}^{\rho}$. Naturally, we are interested in the symmetric portion of it. We have shown that the connection (\ref{con1}) is metric compatible.
 In first order variables it is given by
\begin{eqnarray}
\Gamma_{\nu\mu}^{\rho} &=\partial_{\nu}{\Lambda_{\mu}}^\alpha {\Sigma_\alpha}^{\rho}
+B^{\alpha}{}_{\nu\beta}{\Lambda_{\mu}}^\beta
{\Sigma_\alpha}^{\rho} + \partial_{\mu}{\Lambda_{\nu}}^\alpha {\Sigma_\alpha}^{\rho}
+B^{\alpha}{}_{\mu\beta}{\Lambda_{\nu}}^\beta
{\Sigma_\alpha}^{\rho}\label{con11}
\end{eqnarray}
Specialising for galilean transformation, we get
\begin{eqnarray}
\Gamma_{\nu\mu}^{\rho} &=\partial_{\nu}{\Lambda_{\mu}}^\alpha {\Sigma_\alpha}^{\rho}
+B^{a}{}_{\nu 0}{\Lambda_{\mu}}^\beta
{\Sigma_a}^{\rho} + \partial_{\mu}{\Lambda_{\nu}}^\alpha {\Sigma_a}^{\rho}
+B^{a}{}_{\mu 0}{\Lambda_{\nu}}^0
{\Sigma_a}^{\rho}
\end{eqnarray}
Taking variation due to local galilean boosts, we find
\begin{eqnarray}
\delta_0\Gamma_{\nu\mu}^{\rho} &= \partial_{\mu}{\delta_0\Lambda_{\nu}}^\alpha {\Sigma_\alpha}^{\rho} +
                           \partial_{\mu}{\Lambda_{\nu}}^\alpha \delta_0{\Sigma_\alpha}^{\rho}
+\delta_0 B^{a}{}_{\mu}{\Lambda_{\nu}}^0 + \left(\mu \to \nu \right)
{\Sigma_a}^{\rho}
\label{conint}
\end{eqnarray}
Now, explicit calculation 
yields
\begin{eqnarray}
\delta_0\Gamma_{\nu\mu}^{\rho} = 0
\end{eqnarray}

So we find that the local galilean transformations do not change the metric structures $(\tau_\mu, h^{\mu\nu})$ of the NC geometry or metric compatible connection. It changes the set $(v^\mu, h_{\mu\nu})$. That the calculations have been done in special coordinates (the adapted coordinates) is of course no demerit, because we present the results in tensor form which transform covariantly under general coordinate transformation. The benefit of the galilean gauge theory is that it provides us the appropriate transformations for the vierbein and the spin connections in the adapted coordinates.
\bigskip
{\section{Physical origin of Milne boosts}} We will now show that the transformations due to local galilean boost are actually identical with the Milne boosts To see this, note that
from (\ref{tem1}) and (\ref{n3})
\begin{equation}
\delta_0v^\mu =\Sigma_b{}^\mu u^b  
\end{equation}
Now we propose the map connecting the galilean boost parameter ($u$) with the Milne boost parameter ($\psi$),
\begin{equation}
u^b =  \Sigma_b{}^\mu\psi_\mu\label{map}
\end{equation}
to get
\begin{equation}
\delta_0 v^\mu =  h^{\mu\nu}\psi_\nu\label{n66}
\end{equation}
where we have used (\ref{spm}). Equation(\ref{n66}) 
 exactly matches with the Milne boost transformation (\ref{n1}). In this way  we can derive the entire set of Milne transformations using the same map, from the corresponding results of the last section. This is done in the following.

We start by computing $\delta h_{\mu\nu}$. Using the definition (\ref{spm2}) we get
\begin{equation}
 \delta h_{\mu\nu} = \delta\Lambda_\mu{}^a \Lambda_\nu{}^a + \Lambda_\mu{}^a \delta\Lambda_\nu{}^a 
\end{equation}
But from (\ref{delth3}), $\delta\Lambda_0{}^a = u^a\Lambda_0{}^0 = {\Sigma_a}^{\rho}\psi_\rho\Lambda_0{}^0$. Then \begin{equation}
\delta h_{00}= -(\tau_0P_0^\rho + \tau_0P_0^\rho)\psi_{\rho}
\end{equation}
Similarly working out the variation of the other components, we get 
 \begin{equation}
\delta h_{\mu\nu}= -(\tau_\mu P_\nu^\rho + \tau_\nu P_\mu^\rho)\psi_{\rho}\label{n6}
\end{equation}
The variation (\ref{n6}) agrees upto first order in $ \psi$ with the assumed variations (\ref{n2}) under Milne boost, as given in the literature \cite{JK,OBN}. 

It may be shown that the transformations preserve the NC algebra upto first order in $\psi$.
Denoting the transformed quantities by 'prime' we can calculate
\begin{equation}
v'^\mu \tau'_\mu = (v^\mu + \delta v^\mu)\tau_\mu = (v^\mu +h^{\mu\nu}\psi_\nu )\tau_\mu =v^{\mu}\tau_{\mu}= 1
\end{equation}
 Also
\begin{eqnarray}
h'_{\mu\nu}v'^\nu &=& (h_{\mu\nu} - (\tau_\mu P_\nu^\rho +\tau_\nu P_\mu^\rho )\psi_\rho) (v^\nu + h^{\nu\rho}\psi_\rho)\nonumber\\
                &=& - (\tau_\mu P_\nu^\rho +\tau_\nu P_\mu^\rho )v^\nu\psi_\rho +h_{\mu\nu} h^{\nu\rho}\psi_\rho+O(\psi^2)\nonumber\\
                &=& 0+O(\psi^2)
\end{eqnarray}
and finally
\begin{eqnarray}
h'_{\mu\nu}h'^{\nu\rho}= {\delta_{\mu}}^\rho - \tau_\mu { v'}^\rho +O(\psi^2)={P'_\mu}^\rho +O(\psi^2)
\end{eqnarray}

A significant result of our study is the variation of the connection. We have shown that under local galilean boosts the symmetrized affine connection does not change. Using the map (\ref{map}) the invariance of the connection under Milne boosts may be easily studied.  Indeed, invariance of the connection (\ref{D})is ensured because we have proved \cite{BMM2} that the symmetric connection can be cast in the form (\ref{con11}). 
So we can conclude that the Milne boost invariance of the NC geometry is the outcome of invariance under local galilean boosts, to first order in transformation parameter. We now extend the equivalence for finite boosts.
\bigskip
\subsection {Finite boosts} In Galilei Newton dynamics boosts are commutative. So variations of quantities which are linear in the vierbeins remain the same for finite boosts also. Thus $\tau_\mu$ and $h^{\mu}$ are unchanged while $v^\mu$ transforms as (\ref{n66}).
Finite transformation of $h_{\mu\nu}$ will now become, on exploiting (\ref{W}),
\begin{eqnarray}
h'_{\nu\rho} &=& \Lambda '_{\nu}{}^{a} \Lambda '_{\rho}{}^{a}\nonumber\\
 &=& (\Lambda _{\nu}{}^{a} - u^a{\Lambda _\nu}^{0}) (\Lambda _{\rho}{}^{a} - u^a{\Lambda _\rho}^{0})\label{f1}
\end{eqnarray}
Using the mappings it is possible to reproduce the desired result (\ref{n2}) quoted in the literature. First note that the transformation
(\ref{W}) holds for finite boost as the variations are linear in $u^a$. Using the identification (\ref{map}), we can write from
(\ref{f1})
 \begin{eqnarray}
h'_{\nu\rho} 
 = \left(\Lambda _{\nu}{}^{a} - \Sigma^{a\lambda}\psi_\lambda{\Lambda _\nu}^{0}\right) \left(\Lambda _{\rho}{}^{a} - \Sigma^{a\lambda}\psi_\lambda{\Lambda _\rho}^{0}\right)\label{f11}
\end{eqnarray}
Expanding and using (\ref{spm2}) we get the standard result (\ref{n2}).

We have proved that our results for infinitesmal boost properly generalizes for finite boost. To complete the scenario we look at the invariance of the connection under finite boosts. Once again the abelian property of galilean boosts comes to our help. The finite boost can be divided into a large number of infinitesmal boosts. So $\delta_0\Gamma = 0$ which we have derived for an infinitesmal boost above, also holds for finite boosts. 


\section{Conclusion} That the NC geometry posseses in the metric version additional boost-like symmetry was discovered much earlier \cite{ K}. This 'so called Milne boosts' was studied extensively from an algebraic approach \cite{DH}. The issue was not studied from a dynamic approach until, in recent applications \cite {SW}, non relativistic field theories coupled to curved spacetime came to the centerstage. There were some puzzles regarding the issue of symmetry in the flat limit and the solution required the assumption of a connection between the galilean boost parameter and the gauge transformation parameter \cite {SW}. Then the following questions arise: is it not possible to couple free massive Schrodinger field (without any gauge field) with gravity? Even if we allow a gauge interaction, does not such identification reduce the symmetry of the system \cite{BGM}?  Abandoning these issues, search for an inbuilt gauge field in the Newton Cartan structure \cite{JEN, JK, OBN} begun, exploiting the Milne boost redundancy. But it raised more questions than it actually answered. The corresponding connection ceases to be Milne invariant. Milne invariance can be restored, but at the expense of gauge invariance!
One problem gets traded with another.

We have analysed such issues pertaining to the Milne boost symmetry in the framework of galilean gauge theory (GGT) advocated by us 
\cite{BMM2, BM4, BMM1, BMM3, AM}. This theory arose from the necessity of coupling Schrodinger field consistently with curved spacetime.
The curved space version of Schrodinger equation was always considered to be difficult to construct because of the lack of a single nondegenerate metric. The problem was solved by a novel approach  of gauging the symmetry. The algorithm developed by us can be used in principle for any generic field. Later on, the connection with the Newton Cartan space time was established \cite{BMM2, RBPMt} that led to the formulation of the GGT \cite{BMM3, BM4}.  In this scenario Milne boosts
are studied here in the framework of GGT. We have proved certain points:

\begin{enumerate}
\item The transformations of metric NC theory under the local galilean boost of the first order variables (the vierbeins and spin connections)
are exactly equivalent to the Milne boost transformations. The equivalence is demonstrated by a universal mapping derived here. The physical origin of Milne boost is thus uncovered. 

It is known that the local Lorentz transformations do not change the metric components of the Riemannian geometry but NC geometry which is the non relativistic limit of the Riemann Cartan geometry allow the Milne transformations, which are the consequences of local galilean boosts. Our analysis reveals the reason explicitly. Distinction between time and space introduces a unique direction for time flow in NC spacetime and the local galilean boosts changes space but not time. Thus the influence of the local galilean boosts is not compensated in the metric structures. Here the connection, in contrast to the relativistic case,  cannot be expressed solely in terms of the metric. There is a remnant term that is expressed in terms of the galilean boosts \cite{BMM2}. 
\item Starting from first principles we have shown that the symmetric metric compatible connection remains invariant under Milne boosts. Here the connection is given in term of the vierbeins and the spin connection. 
 Also it has been proved earlier that the expression of the said connection is the same as the Dautcourt connection (see equation (\ref{D})). Our analysis thus ensures the invariance of torsionless form of the connection.  

\item From an algebraic approach alternative transformations can be proposed for Milne boosts where the degenerate metric $\tau_\mu, h^{\mu\nu}$ are not invariant, whereas the auxiliary metric elements $(v^\mu, h_{\mu\nu})$ are. It has been reported \cite{JK} that these transformations do not follow
from a reduction of relativistic theories to nonrelativistic theories
limit. We show here that it is not accidental. Working out the Milne boost transformations from the consequences of local galilean boosts
we have shown that those set of transformations where the degenerate metrics $\tau_\mu, h^{\mu\nu}$  remain unchanged, are the unique form of the Milne transformations.

\item In our analysis Milne boost is clearly generated by the local galilean generators. Thus there is no need for any gauge field in the NC geometry. 
\item One may wonder whether the absence of a gauge field would obstruct the obtention of Newtonian gravity in the present formulation. This is not so. Since we have derived the basic elements of the Newton Cartan geometry (for more details, see \cite{BMM2}), we can just follow the standard steps as outlined in the classic text \cite{MTW} to get Newtonian gravity. This derivation does not require the presence of any gauge field.\end{enumerate}
We hope that our results will give a new perspective on Milne boosts,
\section{Acknowledgement}
 One of the authors (PM) thanks the S. N. Bose National Centre for Basic Sciences for the short term visiting associate during which this work was completed.

\end{document}